\documentclass{article}

\usepackage{arxiv}
\usepackage{float}
\usepackage[utf8]{inputenc} % allow utf-8 input
\usepackage[T1]{fontenc}    % use 8-bit T1 fonts
\usepackage{hyperref}       % hyperlinks
\usepackage{url}            % simple URL typesetting
\usepackage{booktabs}       % professional-quality tables
\usepackage{amsfonts}       % blackboard math symbols
\usepackage{nicefrac}       % compact symbols for 1/2, etc.
\usepackage{microtype}      % microtypography
\usepackage{lipsum}
\usepackage{graphicx}
\usepackage{subfigure}
\usepackage{verbatim}
\usepackage{threeparttable}
\usepackage{amsmath,amsfonts,bm}

\def\eqref#1{equation~\ref{#1}}
\def\1{\bm{1}}

\def\vx{{\bm{x}}}

\DeclareMathAlphabet{\mathsfit}{\encodingdefault}{\sfdefault}{m}{sl}
\SetMathAlphabet{\mathsfit}{bold}{\encodingdefault}{\sfdefault}{bx}{n}

\newcommand{\normltwo}{L^2}

\title{A Model-driven and Data-driven Fusion Framework for Accurate Air Quality Prediction}

\author{
 Chunbo Luo\thanks{Corresponding author}\\
  School of Information and Communication Engineering\\
  University of Electronic Science and Technology of China\\
  Chengdu, China \\
  \texttt{c.luo@uestc.edu.cn} \\
  \And
 Haolin Fei $^{1}$\\
  Glasgow College \\
  University of Electronic Science and Technology of China\\
  Chengdu, China \\
  \texttt{feihaolin@std.uestc.edu.cn} \\
   \And
 Xiaofeng Wu $^{1}$\\
  Glasgow College \\
  University of Electronic Science and Technology of China\\
  Chengdu, China \\
\texttt{wuxiaofeng@std.uestc.edu.cn} \\
}

\begin{document}
\maketitle

\begin{abstract}
Air quality is closely related to public health. Health issues such as cardiovascular diseases and respiratory diseases, may have connection with long exposure to highly polluted environment. Therefore, accurate air quality forecasts is extremely important to those who are vulnerable. To estimate the variation of several air pollution concentrations, previous researchers used various approaches, such as the Community Multiscale Air Quality model (CMAQ) or neural networks. Although CMAQ model considers a coverage of the historic air pollution data and meteorological variables, extra bias are introduced due to additional adjustment.  In this paper, instead of using entire data-driven and end-to-end machine learning strategy to replace the further correction procedure, a combination of model-based strategy and data-driven method namely the physical-temporal collection(PTC) model is proposed, aiming to fix the systematic error that traditional models deliver. In the data-driven part, the first components are the temporal pattern and the weather pattern to measure important features that contribute to the prediction performance. The less relevant input variables will be removed to eliminate negative weights in network training. Then, we deploy a long-short-term-memory (LSTM) to fetch the preliminary results, which will be further corrected by a neural network (NN) involving the meteorological index as well as other pollutants concentrations. The data-set we applied for forecasting is from January 1st, 2016 to December 31st, 2016. According to the results, our PTC achieves an excellent performance compared with the baseline model(CMAQ prediction, GRU, DNN and etc.). This joint model-based data-driven method for air quality prediction can be easily deployed on stations without extra adjustment, providing results with high-time-resolution information for vulnerable members to prevent heavy air pollution ahead.
\end{abstract}

% keywords can be removed
\keywords{CMAQ \and LSTM \and Deep Neural Network \and XGBoost \and Air Quality Prediction}

\section{Introduction}

There is an increasing concern of air particle pollution in recent years since various human chronic disease caused by these molecules contaminant, including SO$_2$ (sulfur dioxide), NO$_2$ (nitrogen dioxide) and NO (nitric oxide)~\cite{lelieveld2015contribution}. Several studies have suggested that the exposure to highly polluted environment will lead to cardiovascular diseases~\cite{miller2007long} and respiratory diseases~\cite{pope2006ischemic}. With the fast development of industry and increasing population, the air pollution tends to be a severe problem in Western China. Chengdu, the biggest city in western China, suffered from haze in January and February for more than 50 days.  Therefore, it is important to develop a more effective warning system, especially in urban area. However, precise pollutant detection technique is hard to achieve due to complex spatial distribution, especially for long-term forecasting application. On the other hand, dynamic pollution detection has been associated with a variety of index, such as climate situation and topographical features. In the last two decades, the Community Multiscale Air Quality model (CMAQ)~\cite{appel2007evaluation,lightstone2017comparing,foley2010incremental} was applied to predict spatial distribution of pollutant in various scales and provided a model for specific pollutants. In addition, Weather Research and
Forecasting Model (WRF) ~\cite{chuang2012erratum}was coupled with CMAQ to include the chemical information in the whole system. However, the CMAQ model is less effective to consider a combination of time scale and spatial distribution, which leads to systematic errors in measurements~\cite{lightstone2017comparing}. In addition, the CMAQ model is limited by its grid prediction feature~\cite{queen2008examining} which is unable to forecast the air conditions in a higher spatial resolution. An improvement mean aims to combine the Atmospheric Dispersion Modelling System (ADMS) in Hongkong ~\cite{fung2018fine}, where further chemical dispersion information of particulate matter is provided. However, the results are still not accurate due to ignorance of long time chemical reactions. Apart from the CMAQ model, the Geographical information systems (GIS)~\cite{briggs2005role} and the nested air-quality prediction modeling system (NAQPMS)~\cite{wang2001nested} are also common models to predict chemical contaminants, but they are not able to handle a large range of input variables due to the relatively limited model capacity. 

Most recently, various data-driven approaches have achieved great advances in constructing linear and non-linear models to predict the trend of pollutants. Primarily in the 1990s,  radical basis functions (RBFs), and generalized regression neural networks (GRNNs) tried to predict the weather condition~\cite{gardner1998artificial,zhou2014hybrid}. To address the time-varying effect, the artificial neural networks (ANN)~\cite{fernando2012forecasting,fu2015prediction,niska2004evolving} has drawn overwhelming attention to approach the complicated distribution of contaminants, e.g., the fusion of simple multi-layer perception (MLP) and Monte Carlo simulations (MCS)~\cite{arhami2013predicting} which take the seasonal factor as a parameter to predict criteria pollutants. Furthermore, various state-of-the-art models presents the relation between air pollutants and meteorological variables such as temperatures, relative humility, and wind speed~\cite{bai2016air,zhou2014hybrid}. The internal relationship between different air pollutants is further put forward to achieve more accurate output~\cite{kumar2017prediction}. However, most of the previous proposed neural networks are unable to construct a long-term memory model, where the existed popular neural networks (MLP, RBFs, NN) may lead to data inconsistency and inaccuracy. To take time information in consideration, a recurrent neural network (RNN) that involves time series is capable of learning time information and building up a sophisticated form of mapping as well. Apart from deep-learning tool, the non-linear machine learning algorithm namely the extreme gradient boosting (XGBoost) ~\cite{chen2016xgboost,zamani2019pm2}which aimed to measure the importance of input factors was applied to predict pollutants concentrations. 

Motivated by the above theory, in this paper, we aim to construct a physical temporal collection (PTC) model to correct the bias from CMAQ predictor and real distribution if adequate historical records are exploited. The ensemble model includes a cascaded LSTM (C-LSTM)~\cite{hochreiter1997long,li2017long,greff2016lstm,zhao2017lstm} to complete time series prediction, and auxiliary information to further correct time series output. For auxiliary information, we investigate the meteorological data, mutual associations between air pollutants and seasonal factors. Moreover, we estimate potential correctness effect by human factor, such as holidays. To better understand the temporal and weather prospective, we propose a temporal pattern and a weather pattern to determine feature importance of input sequences by XGBoost and we remove less important features under threshold value. After eliminating the negative effect induced by irrelevant nodes, the patterns learn to capture important features and improve the model performance. In our study, we also conduct both long-term forecasting and short-term forecasting to explore the temporal dependency of the proposed model. The forecasting process consists of 4 parts: (1) The CMAQ model is applied as a prior predictor for model training on a different time scale. (2) The temporal pattern and weather pattern are employed to capture important features and eliminate negative input variables. (3) The C-LSTM utilizes the CMAQ prediction values and previous air monitoring index to learn the temporal properties and lead air quality prediction. (4) The results from the C-LSTM are further corrected by a deep neural network (DNN) that involves the auxiliary information. 

Our main contribution can be summarised as follows:
\begin{itemize}
\item [1)] We develop a data-driven model to involve DNN and LSTM model utilizing the historical pollutants concentrations and auxiliary data. The original CMAQ correction approaches such as ADMS which only leads to short-term prediction are replaced by deep-learning framework. Compared with the previous baseline (CMAQ prediction), the proposed model improves the precision and suppress the extreme prediction value. 
\item [2)] We adopt XGBoost to construct both temporal pattern and weather patter to measure the prominent factors of input variables and remove the insignificant nodes. We also conduct both short-term forecasting and long-term forecasting and demonstrated that the results outperform the other in different time scales.
\item [3)] We evaluate the robustness of our proposed method by applying the measured value in some major cities that suffer from air quality problems (London), verifying the correctness of both data-driven structure and model-driven structure. The fusion of these two structures overcome their own drawback respectively. 
\end{itemize}
The rest part of the paper is organized as follows. In Section 2, we will introduce 4 different model strategies and data including various air pollutants and climate index. In Section 3, we compare our results with some previous works, such as traditional neural network or linear regression model. Future improvement and conclusion will be covered in Section 4 and Section 5, respectively.

\section{Methods and Data}

\subsection{CMAQ Model}
\begin{figure}[H]
\centering
\includegraphics[width=9cm]{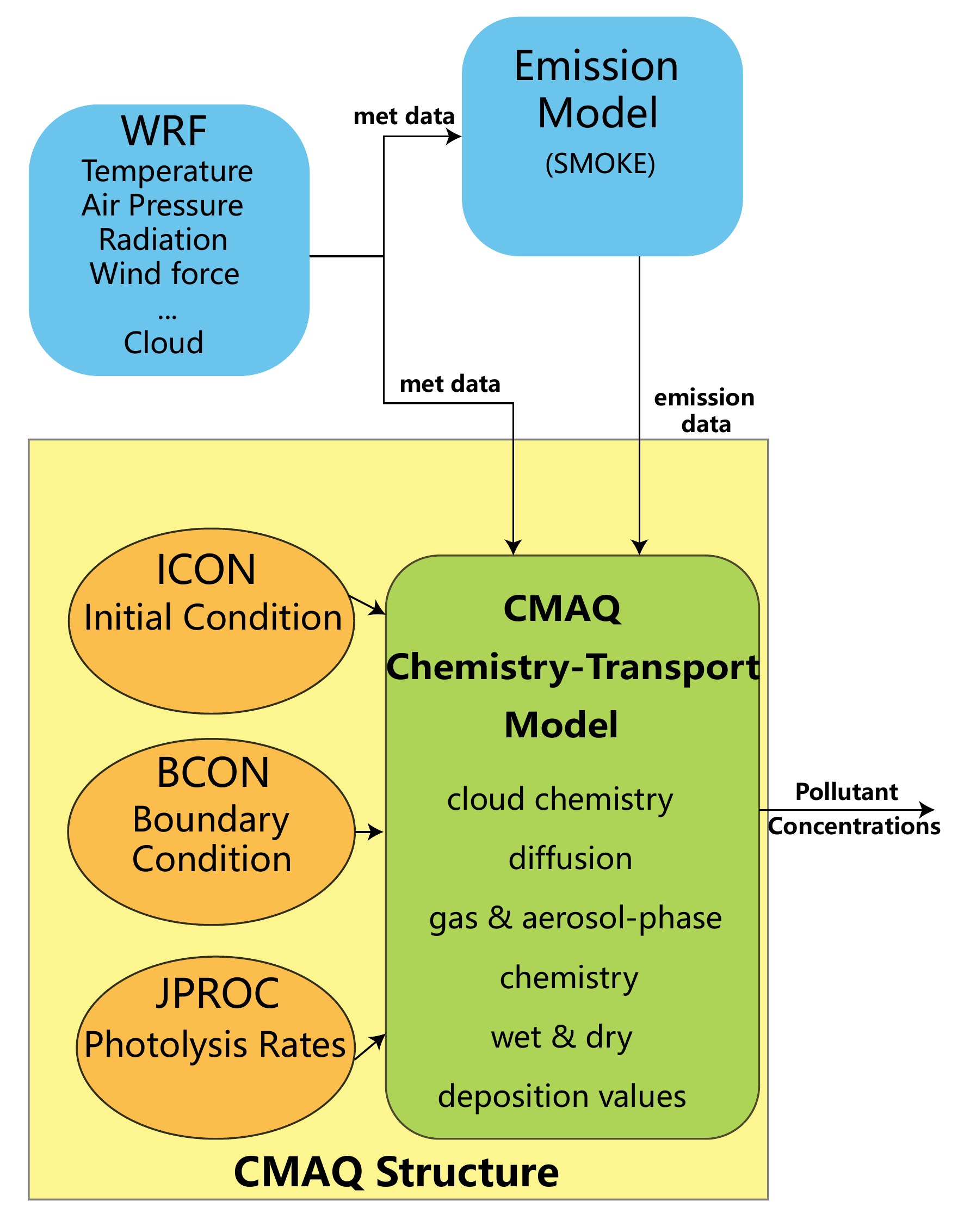}
\caption{The CMAQ Modeling Architecture}
\label{CMAQ_arc}
\end{figure}
As the increasing concern of the air quality issue in these years, the CMAQ was applied to regulate the environment and protect the vulnerable people from 2000. In this literature, we used CMAQv5.1 to predict all the pollutants in 2016. Fihure\ref{CMAQ_arc} shows the modeling architecture of the CMAQ predicted system which includes 5 main components: ICON, BCON, JPROC, MCIP, and CCTM. To determine the simulation boundary condition, ICON and BCON establish  the initial conditions and horizontal boundary respectively. In other words, these two blocks define either spatial condition (36-km, 12-km and 4-km) and temporal condition (7 days) in CMAQ field. When the initial condition is generated, JPROC provides options to select different chemical mechanism to calculate photolysis rates. Moreover, the meteorological model (WRFv3.7) are coupled with CMAQ model to transfer meteorological index into the emission model and CMAQ. The emission model depends on a Sparse Matrix Operator Kernel Emissions (SMOKE) to provide the pollutant sources as input values in CMAQ. In CCTM, the previous inputs and settings are utilized to simulate some chemical conditions as well as calculate some variables, e.g.wet and dry deposition. A hourly prediction is conducted to predict the resulting pollution concentrations.

\subsection{Deep learning algorithm}
This paper propose a  model that combines both traditional model and deep-learning algorithm. A data-driven model regulates the output of traditional model to adjust the prediction value.   In~\cite{lightstone2017comparing}, CMAQ-NN was proposed and it required fewer input data which can be applied in areas with great air quality or slight climate changes. However, for more complex environment such as coastal city, the rapid change in climate cannot been captured properly. Therefore, CMAQ-NN is not suitable for such condition and a more complicated network is required to capture the varying features. We propose a Physical Temporal Collection (PTC) whose input have 3 sources:  CMAQ prediction(24-h CMAQ prediction, 48-h CMAQ prediction, 72-h CMAQ prediction), previous pollutants detection, and meteorological data.

Based on the temporal correlation of air pollution prediction problem, the first model was proposed to find sequential characteristics of the air pollutants. There are several models that can be used for time series prediction task. These models can be concluded roughly as traditional sequential models and state-of art models. For traditional methods, models like auto-regressive (AR), moving average (MA) and auto-regressive moving average (ARMA) are used frequently. For more advanced methods, RNN, convolutional neural network (CNN), and extreme gradient boosting (XGBoost) are used to handle time series prediction tasks in recent years, where RNN is a typical deep learning sequential model. RNN includes two mainstream variations: gradient recurrent unit (GRU) , and LSTM, which we used as our first model.

\subsubsection{LSTM model}

\begin{figure} [H]
    \centering
        \includegraphics[width=1.0\linewidth]{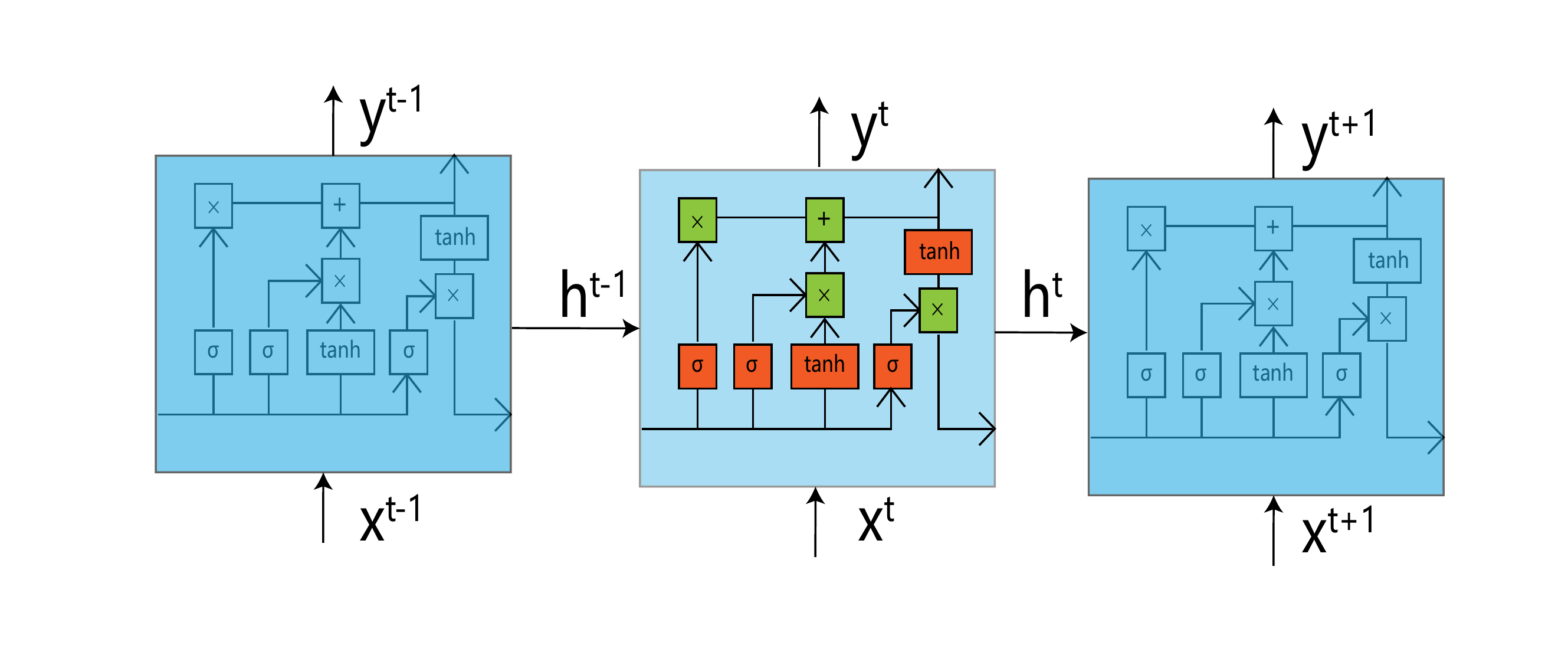}
	  \caption{The LSTM cell}
	  \label{LSTM_neuron} 
	\end{figure}
	
As one of the most popular recurrent neural network, LSTM learns to deal with large-scale data processing compared with other common RNN model, such as GRU. In 1997, the LSTM was first proposed  by Schmidhuber to determine both long-term and short-term dynamic contracts in time series. Unlike the previous RNNs, LSTM take the advantage of constructing long temporal dependencies by its memory ability. Consider that we intend to predict continuous air pollutants based time series inputs, long time dependency of LSTM becomes an efficient to lead accurate results. Moreover, the LSTM requires no calculation according to the parameter of the data-set, which is computationally economical to our problem. In The architecture of LSTM could be represented in Figure. In the picture, each cell block contains three gates, output response and cell state. In first step, the forgot gate $f_t$ will decide whether the information should be saved and transfer the output to the cell state. The calculation of forgot gate $f_t$ is illustrated as:

\begin{equation}
f_{t}=\sigma\left(W_{f} \cdot\left[h_{t-1}, x_{t}\right]+b_{f}\right)
\end{equation}

where $\sigma$ specifies the sigmoid function, $h_{t-1}$ and $x_t$ are the previous LSTM output and current input. $W_f$ and $b_f$ are weights and bias respectively. Input gate $i_t$ has the similar form of forgot gate $f_t$, which can be represented as:

\begin{equation}
i_{t}=\sigma\left(W_{i} \cdot\left[h_{t-1}, x_{t}\right]+b_{i}\right)
\end{equation}

where $W_i$ and $b_i$ are weight and bias of input gate. Then, the calculation of candidate value $\widetilde{C_t}$ as the combination of previous LSTM output $h_{t-1}$ and current input $x_t$:

\begin{equation}
\tilde{C}_{t}=\tanh \left(W_{C} \cdot\left[h_{t-1}, x_{t}\right]+b_{C}\right)
\end{equation}

Then, the update of cell state via prior gate outputs can be yielded as:

\begin{equation}
C_{t}=f_{t} * C_{t-1}+i_{t} * \tilde{C}_{t}
\end{equation}

After the cell state is updated, we can draw the current prediction output $o_t$ as follows:

\begin{equation}
o_{t}=\sigma\left(W_{o}\left[h_{t-1}, x_{t}\right]+b_{o}\right)
\end{equation}

For the C-LSTM, the outputs from the first LSTM model are regarded as the input variables in our second LSTM. In ~\cite{zhao2017lstm}, experimental results presented the C-LSTM outperformed non-cascaded LSTM model by several aspects, including accelerate convergence speed and exclude over-fitting. Therefore, we adopt a C-LSTM here to help the training process.

\subsection{Extreme Gradient Boosting model}
\begin{figure}[H]
    \centering
    \begin{minipage}[c]{0.5\linewidth}
        \centering
        \includegraphics[width=6.5cm]{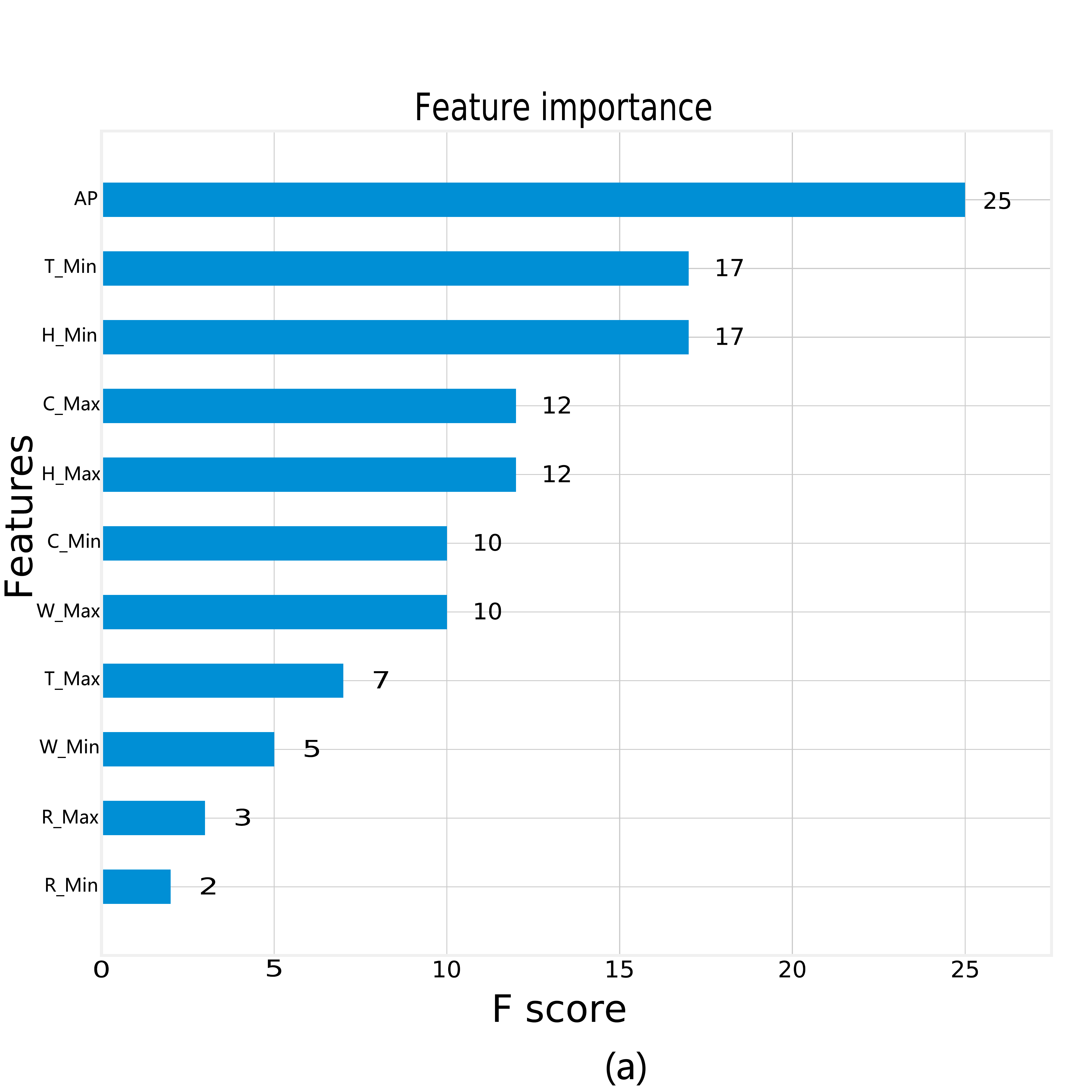}
    \end{minipage}%
    \begin{minipage}[c]{0.5\linewidth}
        \centering
        \includegraphics[width=6.5cm]{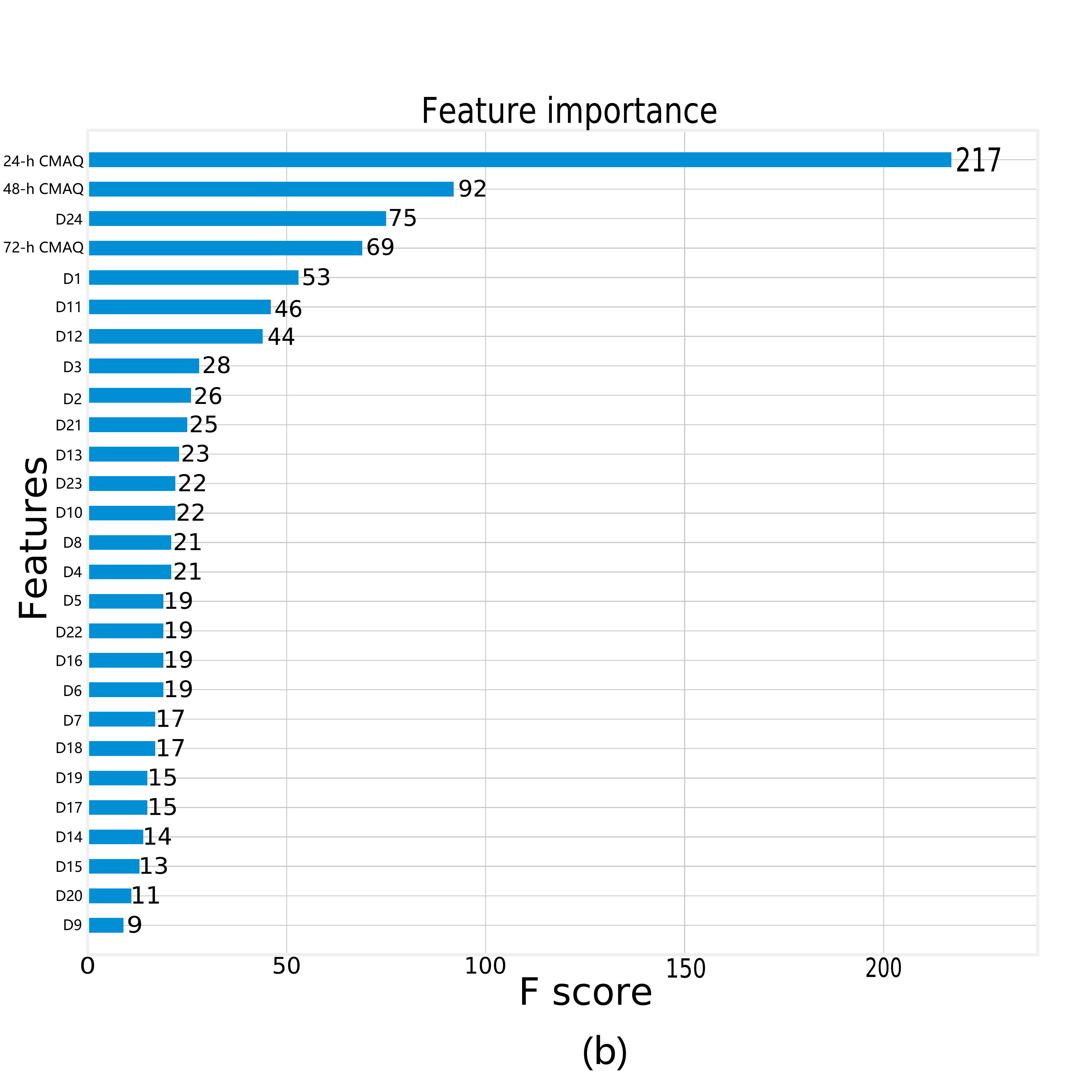}
    \end{minipage}
    \caption{Importance assessment of (a) weather pattern (b) temporal pattern}
	 \label{feature} 
\end{figure}

There are too many potential temporal factors and auxiliary factors. It is important to learn more important feature weights between air pollution and related factors. Extreme Gradient Boosting (XGBoost) is the development of Gradient Boosting model, whereas its objective function is optimized in training process and it has high achievement in both regression and classification problems. The feature will have higher score if it is more used in building the decision tree. Based on the measured score, we set threshold value for both LSTM and DNN input variables to remove the interference factors in data-set. Therefore, the reserved features learn more important weight in network training. In this section, we utilize XGBoost Algorithm to calculate the feature score in training model. All of the affected factors are estimated in proposed model. The basic principle of XGBoost can be represented as follows,
\begin{equation}
{Y_{j}=\sum_{i=1}^{K} f_{i}(X)}
\end{equation}
where $f_{i}(x)$ denotes weak learners that calculate the leaf weights and $Y_{j}$ represents the optimized output of one feature. The model has K classification and regression trees (CARTs) and the feature scores are calculated by summing the squared error of loss function from each CART independently. In each CART, the  feature calculation can be determined as,
\begin{equation}
{y_{j}=\sum_{m=1}^{L} f_{i}(x_m)}
\end{equation}
where the CART has $L$ nodes and $y_j$ represents feature importance in one regression tree. For each inner node $m$, the partition process splits the node to subregion of the whole field. The score for separate feature is measured by optimizing the regularized objective function in iterative process. As discussed before, we employ various input variables into XGBoost for air quality prediction. For the temporal pattern, the input sequences can be defined as (D1,D2,...,D24\footnote{Abbreviate as: D1 (input variable in the first detection period), D24 (input variable in the last detection period)}, 24-h CMAQ, 48-h CMAQ, 72-h CMAQ), and the importance assessment is illustrated in Figure \ref{feature}(b). Moreover, the weather pattern is established by evaluating the essential factors in auxiliary information (AP, T, H, W, R\footnote{Abbreviate as: Atmosphere pressure (AP), Temperature (T), Humility (H), Wind speed (W), Rainfall amount (R)}), which is represented in Figure \ref{feature}(a). The importance assessment analysis is consistent with climate experimental model. For example, ~\cite{huang2017interaction} proves that the low air pressure and high temperature has close correlation with serious air pollutants, such as PM2.5, NO$_2$, SO$_2$ on the subsequent day. Meanwhile, for the temporal pattern, the CMAQ prediction in different time scale and the temporal factor at the same time point in previous days are regarded as the most important features in LSTM training process. 

For hyperparameter selection, the best results for XGBoost were obtained using the following terms: a learning rate of 0.05, a maximum depth of 5, a minimum child weight of 5 and 500 base learners.

\subsubsection{PTC architecture and parameter selection}

\begin{figure} [H]
    \centering
        \includegraphics[width=1.0\linewidth]{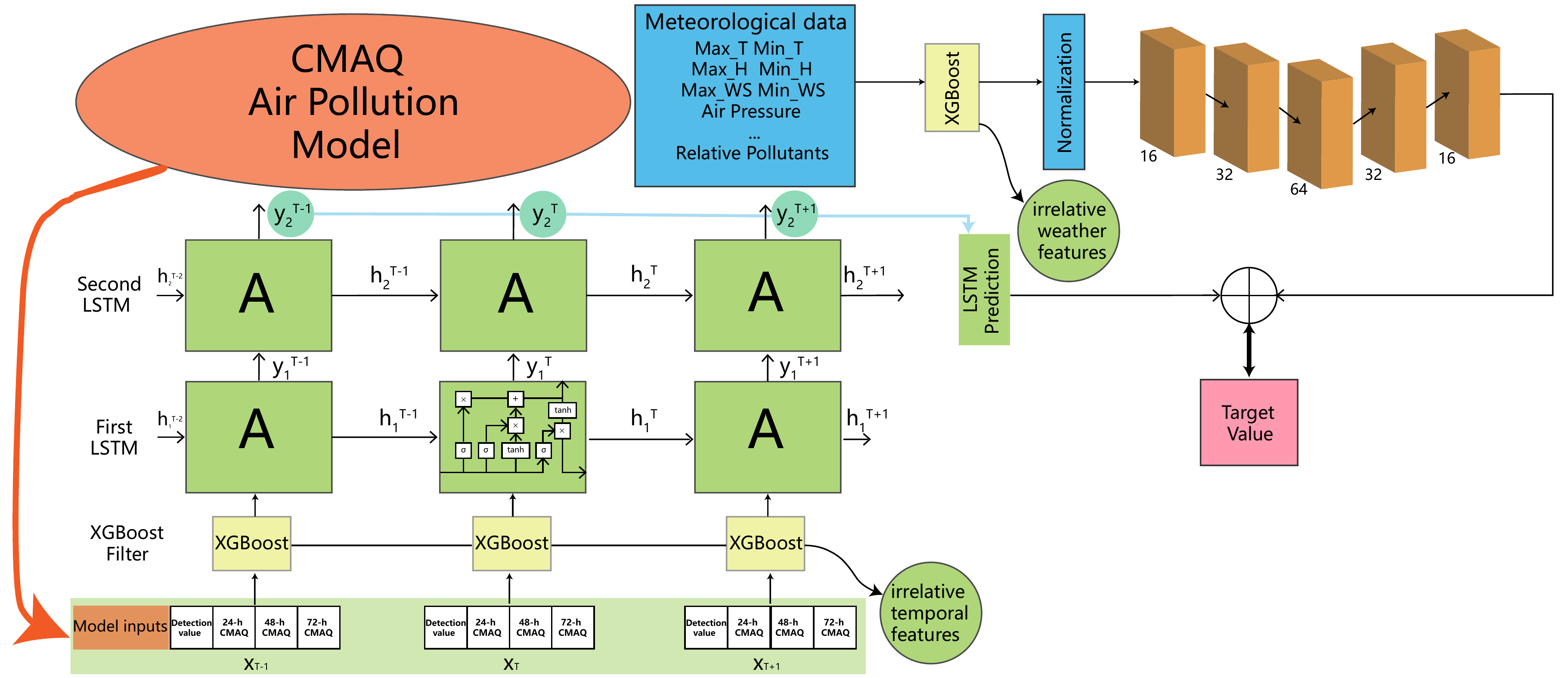}
	  \caption{The structure of PTC.}
	  \label{model} 
	\end{figure}

The architecture of PTC is given in  Figure \ref{model}. There are two LSTM layers cascaded after CMAQ model. For each LSTM layer, there is a dropout layer to avoid over-fitting. The first LSTM model has 50 nodes,with no sequences returning. The second one has 100 nodes and return sequences. The first and the second dropout rate is set to 20\%. A dense layer using linear activation function is used to output values. The network is trained via “adam” optimizer with 0.01 learning rate and batch size is set to 128 and the total loss function use "mse" as penalty function.

Then, we propose a bias correcting technique which takes meteorological data as the input to a deep neural network, and use these data to further approximate the real distribution. With the help of this meteorological feature extracting model, some important features dismissed by CMAQ model or C-LSTM can have more weights, and therefore, affect the prediction values. The input variables include temperature, humidity, wind speed, and air pressure. This bias correction technique calculates the weight metric by a neural network and determine the mutual influence of these contracts. For the temporal pattern, the input sequence is defined as $D_1, D_2, ..., D_{24}$, 24-h CMAQ, 48-h CMAQ, 72-h CMAQ. The weather pattern is also applied to measure important factors of auxiliary information,

The first prediction value form C-LSTM is then compared to the real value which gives a new error array. Meteorological data is then added to the deep neural network to minimize this error array. The first, second and the third dense layer is followed by a dropout layer with 20\% dropout value. Each hidden layer has 16, 32, 64, 32, 16 nodes respectively, and use "relu" activation function for all five hidden layers. The output layer use "sigmoid" activation function with one node. After predicting using "adam" optimizer, the result is then added to the prediction value from our last model to form our final prediction which is weather-pattern sensitive.

\subsubsection{Data processing}
CMAQ prediction values and real observed data are processed, and transform from time series to trainable input and output sequential pairs. First, data are scaled using min-max normalization to mapping values between 0 and 1. The mapping function is shown below:

\begin{equation}
   x^{*}=\frac{x-\min }{\max -\min }
\end{equation}

where $max$ stands for the maximum value of the data, $min$ stands for the minimum value of the data, and $x$ and $x^*$ stands for the array being transformed before and after. The meteorological data used in the second part of PTC model using the same procedure to eliminate the influence of dimension. All the scaled data are transformed inversely after training. 80\% of data are split for training and 20\% are split for testing.

\section{Results and Discussion}

\subsection{Research Area and Experimental Data}
\begin{figure}[H]
\centering
\includegraphics[height=8.5cm]{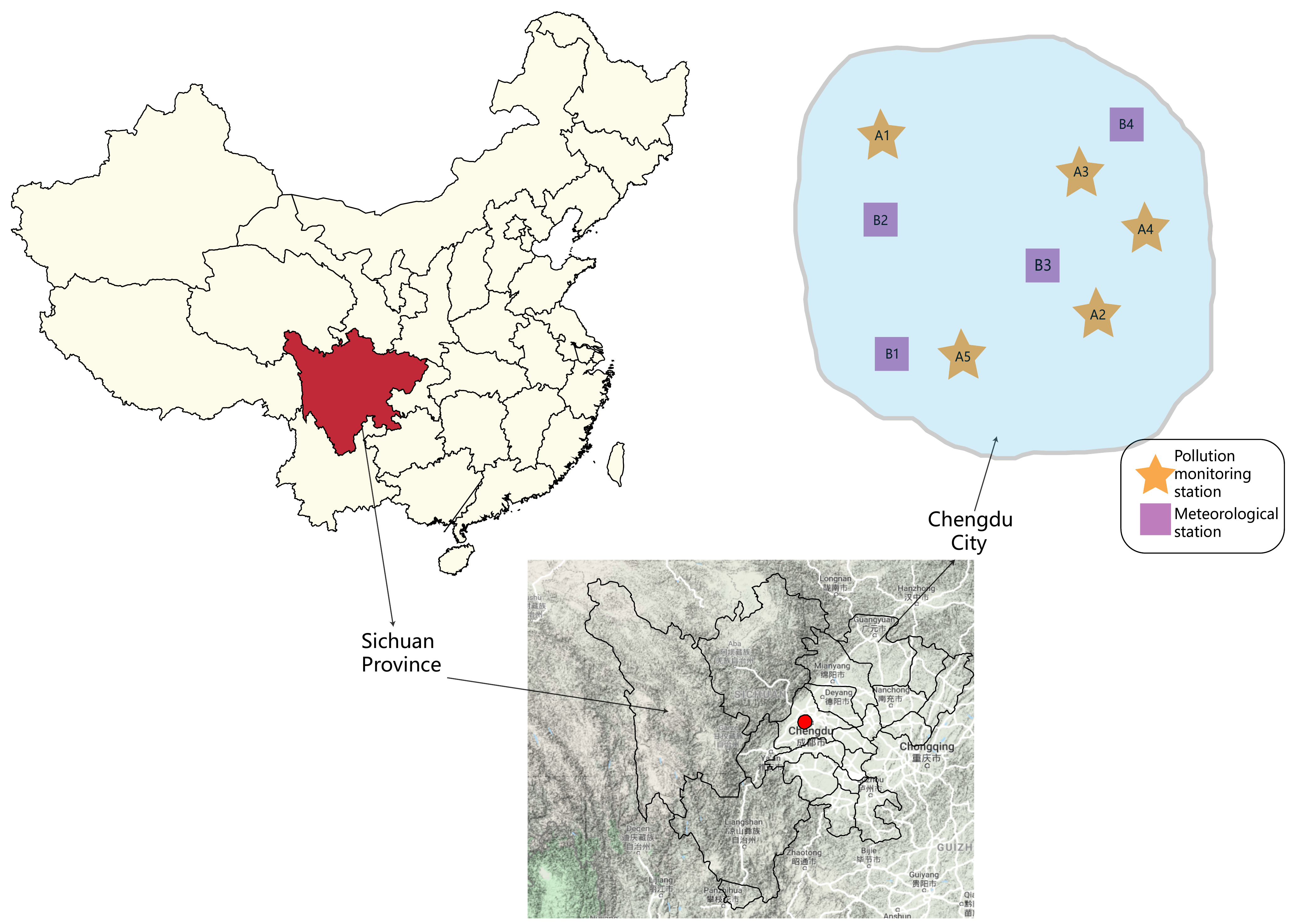}
\caption{Location of the air monitoring stations and meteorological stations in urban area of Chengdu, Sichuan, China. Station A1 to A5 are five main pollution monitoring stations distributed in Chengdu. B1 to B4 are 4 ground meteorological stations distributed in Chengdu.}
\label{chengdu_map}
\end{figure}

With the fast development of the economy and industry, air pollution becomes a challenge to Chengdu. According to the data from World Health Organization, Chengdu was continuously suffered from increasing AQI from 2010 to 2018. Besides, the population of the city reached 16.33 million in 2018. The citizens have high possibility to be exposed to air pollutants. Therefore, constructing urban-wise and long-term air pollution system becomes significant to better control the pollutants and protect people's health. The Figure\ref{chengdu_map} indicates the distribution of air pollution stations and meteorological stations, which can cover the main urban area in Chengdu\footnote{The main urban area of Chengdu is approximately defined by its First Belt Highway} to provide pollutants index. The 5 air quality stations are capable of evaluating the total pollution condition in Chengdu, namely, "Jingquanlianghe", "Liangjiaxiang", "Shilidian", "Sanwayao", and "Shahepu". To be brief, we use the following code in this paper to refer to these stations based on their locations: "Jinquanlianghe" refers to A1, ,  "Shahepu" refers to A2, ”Liangjiaxiang” refers to A3, ”Shilidian” refers to A4 , and ”Sanwayao” refers to A5.

All air quality data are collected from the Environmental Protection and Research Institute of Chengdu\footnote{\url{ https://www.cmascenter.org/cmaq/}}. The meteorological data are available at the National Meteorological Information Center (NMICC), China\footnote{\url{http://data.cma.cn/}}.The data collected and used in this paper is corresponding to a subtropical humid monsoon climate, with high frequency static wind, significant urban heat island effect in summer and stable atmospheric stratification in winter. These characteristics are typical in low speed perennial wind or in basin terrain. The pollution type in Chengdu is coal-smoke air pollution~\cite{Chengdu2014}, and monitoring stations are situated mostly in the urban area, which suggest this data set can represent a industrial metropolitan city model. The pollutant data-set includes most of common pollutants, including CO, NO$_2$, SO$_2$, ozone one-hour average (O$_3 1h$), ozone eight-hour average (O$_3 8h$). The meteorological bias correcting technique is evaluated using meteorological data, which includes the following terms: maximum temperature (max\_T), minimum temperature (min\_T), maximum humidity (max\_H), minimum humidity (min\_H), maximum wind speed (Max\_WS), minimum wind speed (Min\_WS) and air pressure from 24 hours ago.

Although meteorological condition changes from year to year, models described in the next few sections, trained from data in 2016, can perform almost the same as it did in 2015 and 2017. Based on that, we take the meteorological data from 1 January 2016 to 30 June 2017 as the training data, and data after July 2017 as the testing and validation data for all methods described above. In addition, for each pollutant, their own models are established instead of using just one model for all the pollutants. For simplicity, “true value” refers to the real value measured by the meteorological value and “CMAQ 24 hour”, “CMAQ 48 hour”, “CMAQ 72 hour” refer to the CMAQ prediction value forecast 24 hours, 48 hours and 72 hours ago. 

\subsection{Model performance evaluation}
We evaluate the performance of PTC by comparing the predicted values with real values via three types of measurements in dataset: the Mean Absolute Error(MAE), the 
Root Mean Squared Error (RMSE), and Coefficient of Determination(R$^2$). 
The MAE is defined as:

\begin{equation}
   \mathrm{MAE}=\frac{1}{\mathrm{m}} \sum_{t=1}^{\mathrm{m}}(\mathrm{y}(t)-\widehat{\mathrm{y}}(t))
      \label{e2}
\end{equation}

The RMSE is defined as:

\begin{equation}
   \mathrm{RMSE}=\sqrt{\frac{1}{m} \sum_{t=1}^{m}\left(\mathrm{y}(t)-\hat{\mathrm{y}}(t)\right)^{2}}
   \label{e1}
\end{equation}

The R-square is defined as:

\begin{equation}
   \mathrm{R}^{2}=1-\frac{ \sum_{t=1}^{\mathrm{m}}(\mathrm{y}(t)-\widehat{\mathrm{y}}(t))^{2}}{ \sum_{t=1}^{\mathrm{m}}(\mathrm{y}(t)-\bar{\mathrm{y}}(t))^{2}}
      \label{e3}
\end{equation}

Moreover, we validate our model by measuring the Euclidean distance between CMAQ 24 hours’ prediction ($\epsilon_{base}$) and the real value together with our bias correction model prediction ($\epsilon_{model}$) and the real value:
\begin{equation}
   \varepsilon_{\text{base}}=\frac{1}{L} \sum_{t=1}^{\mathrm{m}}\left\|Y_{\text{CMAQ} 24 \text{h}}-Y_{\text{true}}\right\|_{2}^{2}
\end{equation}
\begin{equation}
    \varepsilon_{\text{model}}=\frac{1}{L} \sum_{t=1}^{\mathrm{m}}\left\|Y_{\text {model }}-Y_{\text {true }}\right\|_{2}^{2}
\end{equation}
where $L$ stands for the number of time point in the test set, Y$_{\rm CMAQ}$ stands for the CMAQ 24 hour prediction value vector of CMAQ output, Y true stands for the real value vector, Y$_{\rm model}$ stands for every model prediction vector and  $|| \vx ||$ is the $\normltwo$ distance of two vector. To compare the extent our model surpass the CMAQ prediction, the following equation is used:
\begin{equation}
    Accuracy=\left(\varepsilon_{\text{base}}-\varepsilon_{\text{model}}\right) / \varepsilon_{\text{base}} \times 100\%\label{con:eq3}
\end{equation}
We use 24 hour CMAQ prediction as our baseline instead of 48 or 72 hours for the reason that CMAQ 24 hour prediction has the least Euclidean loss to the real value normally. 

\subsection{Related Model Comparison}
To show the robustness of our PTC model, we also conduct the air quality prediction over the following advanced regression models.
\begin{itemize}
\item [1)] Gated Recurrent Unit (GRUs) + XGBoost: GRU is a simple form of LSTM where the current output y is determined by the time series input x and the hidden state from previous node. The XGBoost constructs the temporal pattern and weather pattern here to select important features of input variables, which is the same as PTC.
\item [2)]LSTM + DNN: LSTM model make continuous prediction based on the temporal input. DNN is applied to involve auxiliary information to improve the prediction accuracy. We remove the XGBoost structure here to demonstrate the superiority of importance assessment process. 
\item [3)]DNN + XGBoost: Deep Neural network are able to make precise prediction in high-dimensional space, which are popular in air pollution prediction. We replace the LSTM by DNN with the whole dataset, including CMAQ prediction values, air quality detection values and auxiliary information. We address that the time series architecture outperform DNN in  air quality prediction problem.
\label{Compared methods}
\end{itemize}

\subsection{Results}

\begin{figure}[h]
\centering\includegraphics[width=11cm]{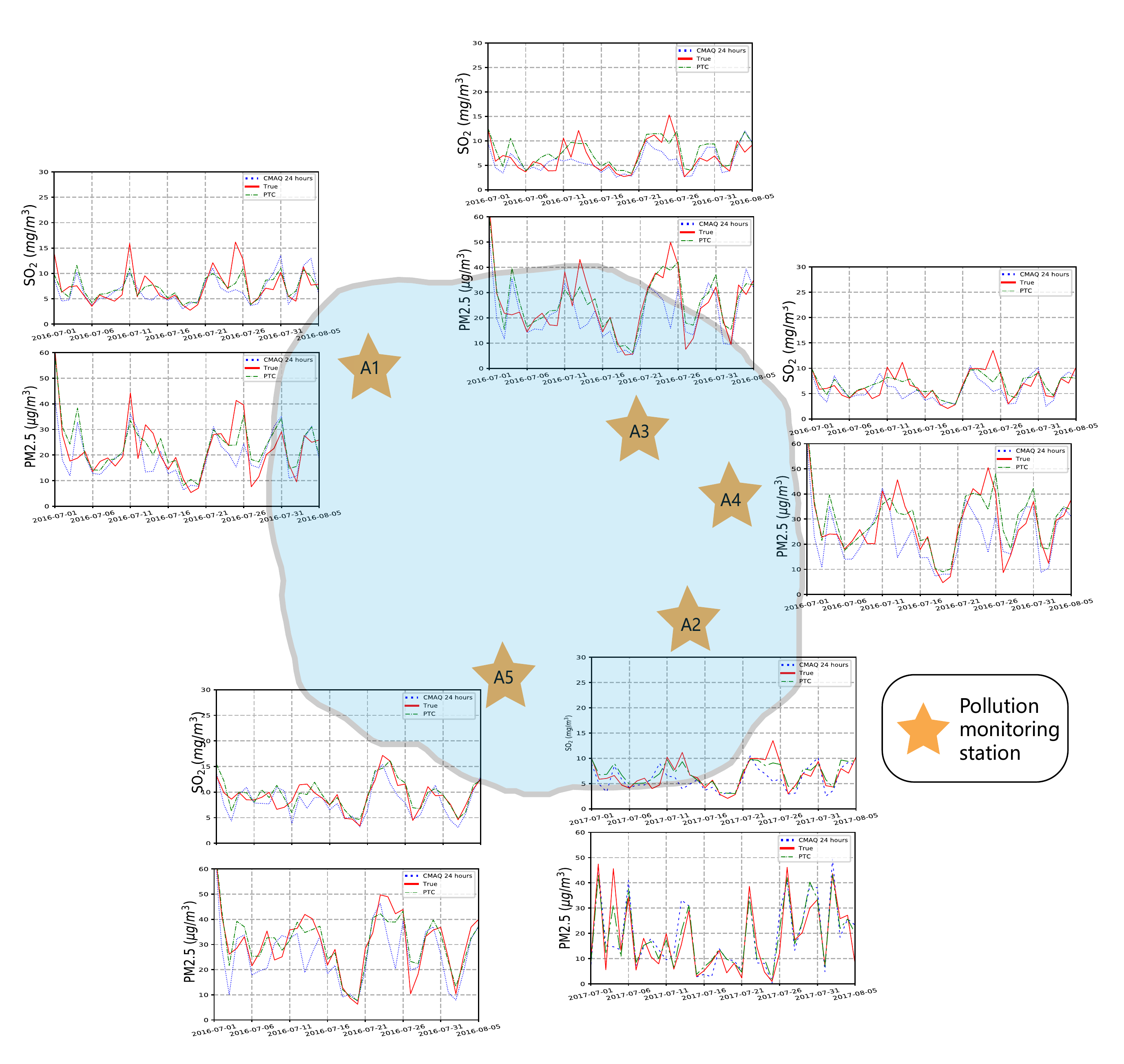}
\caption{Prediction of PM2.5 and SO$_2$ in urban-wise area of Chengdu city}
\label{fig2}
\end{figure}

In this section, we evaluate our proposed PTC model over several aspects. First, we assess the pollutants distribution of SO$_2$ and PM2.5, which are inaccurately predicted by CMAQ in around one month time interval. Secondly, the temporal stability of our PTC is conducted over previous works in terms of MAE, RMSE, and $R^{2}$ for multiple concerning pollutant concentrations. Moreover, we evaluate the temporal dependencies in two categories: long-term forecasting and short-term forecasting in all air monitoring stations. To show the merits of our PTC model, we compare the accuracy improvement using  Equation \ref{con:eq3} with various popular methodologies, including GRU, XGBoost and DNN. All the results are properly shown in figure or table. 

In Figure \ref{fig2}, the prediction values of proposed model at urban-wise Chengdu are shown. The selection period is summer holiday where the air pollutants are under dynamic change. It can be seen from the picture that there exist obvious difference between each air monitoring stations. Therefore, to extract unique feature in each air quality stations, we separately build various PTC model to make precise prediction. At the same time, we presented great suppression effect on extreme prediction values of PTC model over CMAQ prediction since large bias will be corrected by L2 loss. 

Table \ref{Quantitative evaluation} shows the quantitative evaluation indexes of four types of concerning contaminants in five air quality monitoring stations. The performance is measured by comparing the predicted and observed pollutant concentrations via Equation \ref{e1}, \ref{e2} and \ref{e3} in 24-h prediction. It can be shown in the table that the predicted results of $PM2.5$ and $NO_2$ have a close relation with observed values. To examine the temporal dependencies of PTC model, we conduct both short-term and long-term (6-h to 72-h time interval) prediction. In Figure \ref{fig3}, the results of both long-term and short-term prediction is presented, where the MAE increased with the longer prediction time. For short-term forecasting, our predictor draw accurate results due to limited pollutant variation in short period. Although long-term prediction has higher dimensional input, the error is still increasing since irrelevant input interference in high dimensional vector. Therefore, importance assessment of input vector by XGBoost is significant, while the negative effect induced by the irrelevant variables will be reduced. 

\renewcommand{\arraystretch}{3.5} %控制行高
\begin{table}[H]
  \centering
  \fontsize{5}{9}\selectfont
  \begin{threeparttable}
    \begin{tabular}{cccc ccc ccc cccc cc}
    \toprule
    \multicolumn{3}{c}{A1}&\multicolumn{3}{c}{ A2}&\multicolumn{3}{c}{ A3}&\multicolumn{3}{c}{ A4}&\multicolumn{3}{c}{ A5}\cr
    \cmidrule(lr){2-4} \cmidrule(lr){5-7} \cmidrule(lr){8-10} \cmidrule(lr){11-13} \cmidrule(lr){14-16}
    &RMSE&MAE&R$^2$&RMSE&MAE&R$^2$&RMSE&MAE&R$^2$&RMSE&MAE&R$^2$&RMSE&MAE&R$^2$\cr
    \midrule
    CO ($m g/m^3$)&0.4759&0.2314&53.39\%&0.7731&0.4246&46.47\%&0.7605&0.4036&50.04\%&0.6593&0.3771&57.87\%&0.5635&0.3158&33.37\%\cr
    SO$_2$ ($\mu g/m^3$)&29.79&22.44&56.57\%&47.83&39.17&56.13\%&45.72&33.28&45.76\%&43.24&33.05&45.86\%&38.34&29.65&42.63\%\cr
    PM2.5 ($\mu g/m^3$)&43.14&30.58&52.60\%&49.46&37.18&46.23\%&52.69&35.07&47.94\%&47.24&33.26&45.86\%&41.40&29.19&50.45\%\cr
    NO$_2$ ($\mu g/m^3$)&20.04&13.59&58.02\%&23.41&16.41&69.38\%&21.60&15.94&49.14\%&21.96&15.93&51.57\%&19.44&14.71&48.25\%\cr
    \bottomrule
    \end{tabular}
      \caption{Quantitative evaluation in five stations}
  \label{Quantitative evaluation}
    \end{threeparttable}
\end{table}

\begin{figure}[h]
\centering\includegraphics[width=14cm]{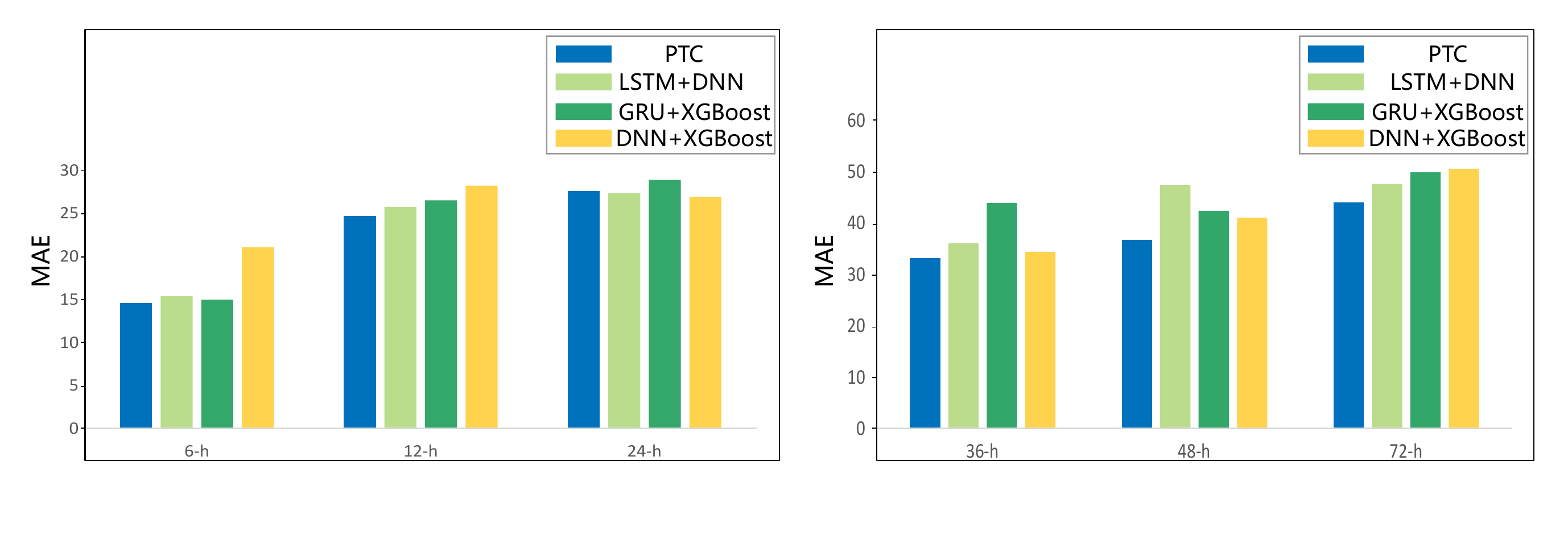}
\caption{Results comparison for short-term and long-term forecasting}
\label{MAE_com}
\end{figure}

The temporal dependencies in short-term and long-term forecasting in all air monitoring stations are given in Figure \ref{MAE_com}, with three previous studies. To keep consistent standards, the compared approaches are mentioned in Section \ref{Compared methods}. From the results, PTC outperforms the other methodologies. It is worth noting that DNN+XGBoost has large prediction bias in most of the time scale. The time series prediction architecture improves forecasting accuracy compared with non-temporal model. PTC are capable of extracting the most important features of inputs via XGBoost, compared with LSTM+DNN method. 

\begin{figure}[h]
\centering\includegraphics[width=14cm]{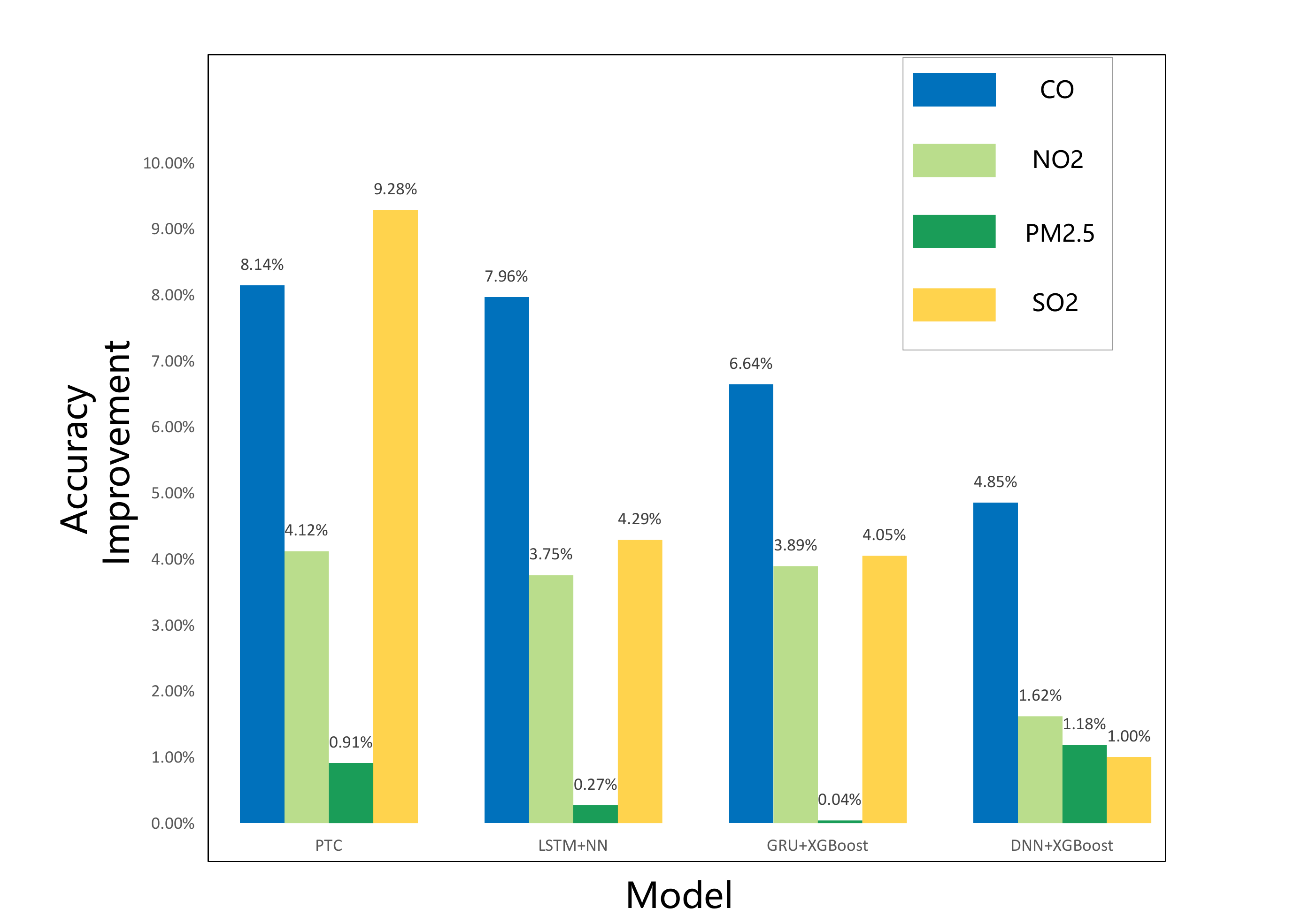}
\caption{Accuracy improvement comparison}
\label{Acc_com}
\end{figure}

\begin{figure}[H]
\centering\includegraphics[width=10cm]{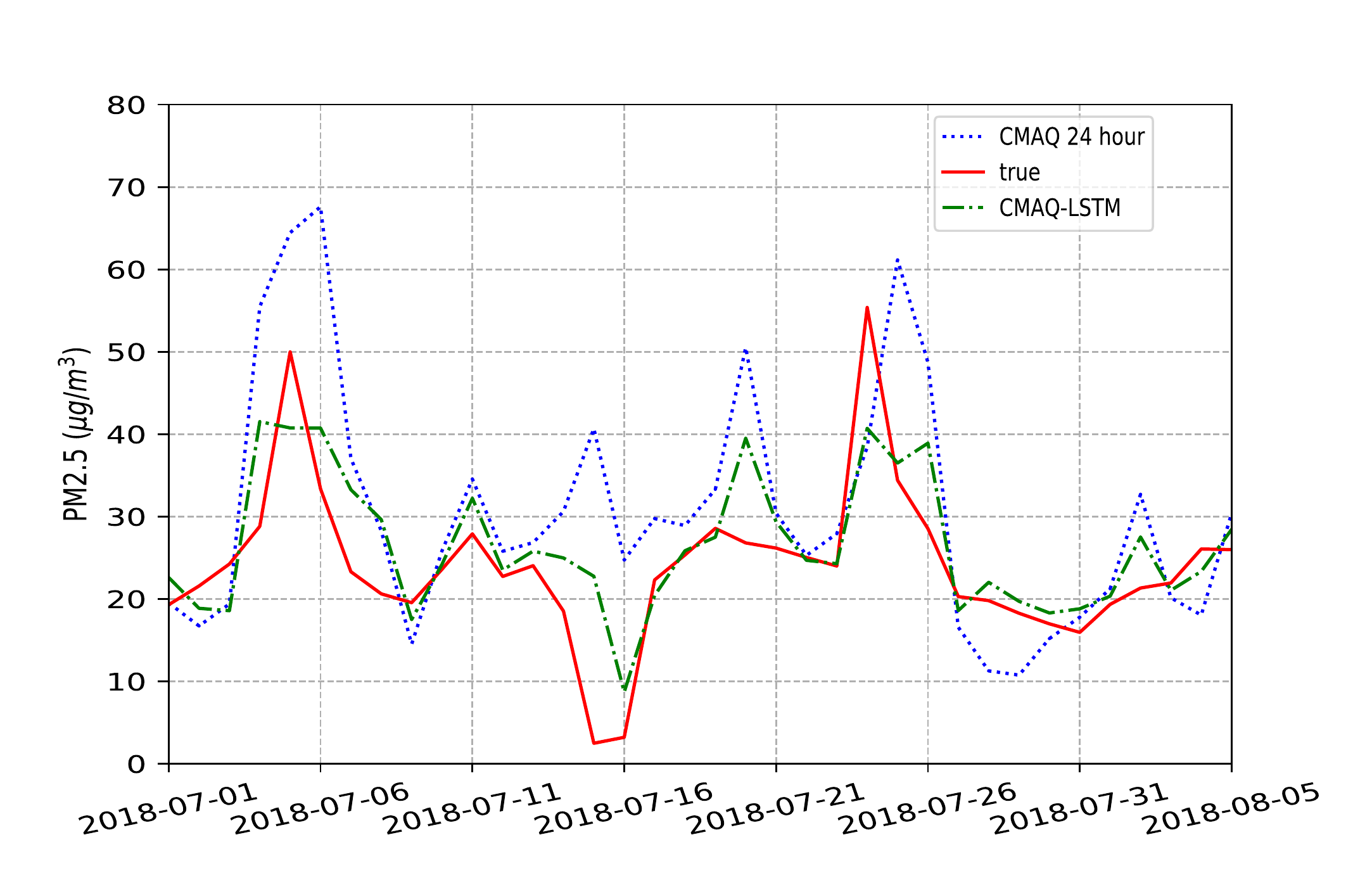}
\caption{The prediction of daily PM2.5 concentrations in Bexley, London}
\label{fig3}
\end{figure}

In Figure \ref{Acc_com}, the five stations averaging accuracy improvement of PTC model over CO, NO$_2$, SO$_2$ and PM2.5 are compared with other approaches. The results reveal that our PTC model has the best performance over all other popular methodologies. However, CMAQ model already has a good prediction accuracy on PM2.5, and the accuracy improvement for PM2.5 is small of all four compared model.

To verify the robustness of our proposed model, we also conduct our PTC structure on London with the air pollutants concentrations in 2018. The air quality data are available to be collected at UK-AIR (Air Information Resource) 4. Only the prediction of PM2.5 is illustrated due to its importance. In quantitative analysis, the proposed model improves the forecasting accuracy by 29.543\%, which validate capability of memorable model to correct the bias from the raw CMAQ prediction.

\section{Conclusions}
In this paper, we proposed a physical temporal ensemble model to give more accurate predictions on pollutants and air quality based on traditional air forecast models. The proposed model utilize the traditional model data, past observed values, and meteorological value to give a more accurate prediction on air pollutants concentration and air quality. This fusion of model-driven and data-driven architecture can cease the systematic error the traditional model has, while the data-driven stage could require less amount of data benefit form the model-driven stage, which used as the first stage of our method. By observing the trend of real air pollutants concentration and extracting meteorological pattern, the CMAQ model prediction accuracy is further improved. In addition, the feature importance, which includes meteorological data and public holidays, are examined to see how much influences these feature are on the performance of CMAQ model.

The assessment on different models shows that our PTC model can produce a more accurate result than using single traditional model or other architecture. It is worth noting that, with the feature importance judgement mechanism using XGBoost, our model can utilized parameters more flexible and efficiently, and develop a precise projection to a time point throughout 24 hours a day. In the future, this PTC can be regarded as a effective model to replace the original ensemble process to achieve more accurate results.

\bibliographystyle{unsrt}  
\bibliography{main}

\appendix
\section{Appendix}
The formula for the ensemble adjustment model is given below:
\begin{equation}
\resizebox{.9\hsize}{!}{$
    {{d}_{1}=1}\times {{\text{(}{{\text{X}}_{\text{train 24h}}}\text{-}{{\text{X}}_{\text{now 24h}}}\text{)}}^{2}}\text{+0}\text{.8}\times {{\text{(}{{\text{X}}_{\text{train 48h}}}\text{-}{{\text{X}}_{\text{now 48h}}}\text{)}}^{2}}\text{+0}\text{.6}\times {{\text{(}{{\text{X}}_{\text{train 72h}}}\text{-}{{\text{X}}_{\text{now 72h}}}\text{)}}^{2}}
    $}
\end{equation}

\begin{equation}
\resizebox{.9\hsize}{!}{$
    {{d}_{2}=0}\text{.3}\times {{\text{((}{{\text{X}}_{\text{train 24h}}}\text{-}{{\text{X}}_{\text{train 48h}}}\text{)-(}{{\text{X}}_{\text{now 24h}}}\text{-}{{\text{X}}_{\text{now 48h}}}\text{))}}^{2}}\text{+}\text{0.2}\times {{\text{((}{{\text{X}}_{\text{train 48h}}}\text{-}{{\text{X}}_{\text{train 72h}}}\text{)-(}{{\text{X}}_{\text{now 48h}}}\text{-}{{\text{X}}_{\text{now 72h}}}\text{))}}^{2}}
    $}
\end{equation}

where X$_{train}$ stands for the training set and X$_now$ stands for the set of value to be validate currently. $d_1$ measures the similarities between the current value and values in the database. $d_2$ measures the similarities between the current trend and trend in the database. The set which has minimum d value means the most similarities between current value set and this set, and the ensemble adjustment value which based on "experience" is directly calculated by looking at how this most similar set modify the predicted value output by traditional model.

\end{document}